# Influence of Ru content on electrocatalytic activity and defect formation of Au-Pd-Pt-Ru compositionally complex solid solution thin films

Miran Joo[1], Huixin Xiu[1,2], Sabrina Baha[3], Ridha Zerdoumi[3,4], Ningyan Cheng[1], Christoph Somsen[5], Yujiao Li[6], Aleksander Kostka[6], Wolfgang Schuhmann[4], Alfred Ludwig[3,6] and Christina Scheu[1*]

[1] Nanoanalytics and Interfaces, Max Planck Institute for Sustainable Materials, 40237 Düsseldorf, Germany

[2] School of Materials and Chemistry, University of Shanghai for Science and Technology, 516 Jungong Road, Shanghai, 200093, China

[3] Chair for Materials Discovery and Interfaces, Institute for Materials, Ruhr University Bochum, 44801 Bochum, Germany

[4] Analytical Chemistry-Center for Electrochemical Sciences (CES), Faculty of Chemistry and Biochemistry, Ruhr University Bochum, 44801 Bochum, Germany

[5] Chair for Materials Science and Engineering, Institute for Materials, Ruhr University Bochum, 44801 Bochum, Germany

[6] Center for Interface-Dominated High Performance Materials (ZGH), Ruhr University Bochum, 44801 Bochum, Germany

*Corresponding author
E-mail address: c.scheu@mpi-susmat.de

# Abstract

Compositionally complex solid solutions (CCSSs) consist of a randomly mixed single phase with the potential to enhance electrocatalytic activity through their polyelemental surface atom arrangements. However, microstructural complexity originating from multiple principal elements influences local structure, chemistry, and lattice strain, which might also affect electrocatalytic activity. Here, we investigate the effect of Ru content on electrochemistry and defect formation in Au-Pd-Pt-Ru CCSS thin films. Such defects could provide active sites when terminating at the CCSS surface or modify surface composition through preferential segregation. A thin-film material library covering a wide composition range was fabricated by room-temperature combinatorial co-sputtering. High-throughput compositional, structural and functional characterization, including electron microscopy equipped with energy dispersive X-ray spectroscopy, X-ray diffraction, and electrochemical screening, were used to correlate composition and microstructural features with catalytic activity. Three representative compositions selected from the library - $Au_{68}Pd_{13}Pt_{15}Ru_{4}$, $Au_{27}Pd_{24}Pt_{23}Ru_{26}$, and $Au_{9}Pd_{21}Pt_{18}Ru_{52}$ - were examined in detail. The three samples exhibit face-centered cubic structures, with lattice contraction occurring with increasing Ru content. In addition, with increasing Ru content, a transition from a high density of nanotwins to high-density, atomic-layer stacking faults was observed. Moreover, the hydrogen evolution reaction activity improves with higher Ru content. Atom probe tomography reveals local compositional fluctuations, including element-specific enrichment and depletion at grain boundaries. The findings provide a new insight into surface atom arrangement design in the CCSS electrocatalysts with enhanced performance.

# 1. Introduction

Compositionally complex solid solution (CCSS) systems—consisting of multiple principal elements in a single-phase solid solution—are of high interest in catalytic materials design[1-6]. While the demand for efficient catalysts continues to grow, Pt-based materials remain limited for large-scale industrial applications due to their high cost and scarcity. To address this, extensive efforts have focused on reducing the Pt content by alloying with transition metals[7-11]. However, the incorporation of less noble alloying elements often compromises phase stability, undermining the long-term durability of electrocatalysts and thereby reducing overall electrocatalytic performance[12,13]. In this regard, CCSS materials offer a promising alternative: by incorporating five or more elements in a highly mixed state, they exhibit high configurational entropy, which supports to stabilize the solid solution state during synthesis but also could support to maintain phase stability under electrochemical environments.

The random elemental mixing of CCSS systems gives rise to diverse surface atom arrangements. These arrangements generate a broad distribution of binding energies across the surface, enabling efficient adsorption and desorption of reaction intermediates and ultimately accelerating reaction kinetics[14-16]. The compositional complexity that provides catalytic versatility can also lead to further properties which could also influence the electrocatalytic activity of these materials. Lattice distortions and severe strain accumulation can occur due to the size differences of constituents[17,18]. To relieve strain, defects such as twins or stacking faults can form. While such features may enhance catalytic activity, long-term stability could pose a critical challenge for CCSS electrocatalysts.

Microstructure introduces additional complexity to CCSS systems. In perfect crystalline solids with long range order atoms are periodically arranged. In reality, however, various defects inevitably may form, ranging from point defects such as vacancies to extended defects such as dislocations and grain boundaries. These features alter local atomic arrangements and electronic structures, thereby also impacting catalytic performance. Moreover, the atom arrangements at surfaces can differ substantially from that in the bulk, especially under electrochemical conditions. Local compositional fluctuations around defect structures are particularly relevant in CCSS systems, where diverse atomic configurations could promote the formation of complex hierarchical microstructures. As a result, defect formation influences catalytic behavior, underscoring the importance of probing structural complexity with advanced characterization methods – a research area that remains underexplored[19-22].

To date, most investigations of CCSS materials have emphasized microstructural analysis in relation to mechanical properties[23-25], while relatively limited studies have addressed the atomic- and nanoscale structures that directly govern catalytic activity. Although CCSS catalysts have begun to attract attention for electrocatalytic applications, much of the research remains at an early stage focusing largely on exploring preliminary correlations between composition, microstructure and electrochemical behavior. Given the extreme compositional and microstructural complexity of CCSS, detailed characterization is needed to unravel the fundamental physics linking atomic-scale features to catalytic properties. Because catalytic processes inherently occur at the nano- and atomic-scales, high-resolution (scanning) transmission electron microscopy (HR-(S)TEM) techniques, in combination with high-throughput characterization and other advanced characterization tools are required to select the most relevant samples from the large compositional ranges of interest. This is especially important for investigating the complexity and establishing the interconnections between composition, structure and activity.

This study investigates the effect of the Ru content on the electrocatalytic properties and defect formation in Au-Pd-Pt-Ru CCSS thin films, with a special focus on planar defects in dependence of the increasing Ru content. High-throughput (screening) methods including scanning electron microscopy (SEM) with energy dispersive X-ray spectroscopy (EDX), X-ray diffraction (XRD), and scanning droplet cell (SDC) measurements allow to correlate the average chemical composition and crystal structure with electrocatalytic activity. On a local scale, HR-(S)TEM analysis reveals grain morphology and composition distribution, and further provides the detailed structural information on nanotwins and stacking faults that develop in the thin films depending on the Ru content. Additionally, atom probe tomography (APT) highlights local atomic fluctuation around grain boundaries, supporting our understanding of the compositional and microstructural complexity in the CCSS thin films.

## 2. Results and discussion

**Figure 1a** shows maps of the compositional gradients of the quaternary thin-film materials library Au–Pd–Pt–Ru system fabricated by magnetron co-sputtering. Each map spans a wide compositional range, with the elemental contents ranging from low to high across the library, corresponding to the positions of the four pure-element sputter targets. With the targets positioned at the sputter chamber corners, composition gradients are achieved across the library, with the equiatomic composition located near the center of the sapphire substrate. High-throughput SEM-EDX measurements were used to determine compositional distributions across the library. The nominal compositions of three representative measurement areas (MAs) are listed in **Table S1 (Supporting Information)**. These values agree closely with low-magnification STEM-EDX measurements **(Table S1)**, confirming the reliability of the compositional analysis.

The electrocatalytic activity of the materials library for the hydrogen evolution reaction (HER) was investigated in 0.05 M KOH using a high-throughput SDC. The linear sweep voltammograms (LSVs), normalized to the geometric surface area, are shown in **Figure 1b**. The potential required to reach a current density of -3 mA/cm² was then determined from the LSVs and plotted as an activity map in **Figure 1c**. The activity map reveals a strong compositional dependence: Ru-rich regions (upper right) exhibit significantly higher activity (lower overpotential), while Ru-poor (Au-rich) regions show a lower activity. The systematic increase in activity with Ru content suggests that higher Ru contents tune the surface energies for hydrogen adsorption by modifying the electronic structure of the active sites with neighboring elements. It is worth noting that the electrochemical measurements for the HER activity were conducted in the presence of atmospheric oxygen, and therefore, the LSVs may contain a contribution from the oxygen reduction reaction (ORR), especially in the low-current regions. Therefore, selecting a relatively high current density of -3 mA/cm², where the Faradaic response is primarily due to the HER, effectively removes the ORR background to a large extent. To establish detailed correlations between composition, structure and activity, three representative compositions (marked with red squares on the activity map) were selected for further investigations: $Au_{68}Pd_{13}Pt_{15}Ru_4$ (Ru-poor), $Au_{27}Pd_{24}Pt_{23}Ru_{26}$ (equiatomic), and $Au_9Pd_{21}Pt_{18}Ru_{52}$ (Ru-rich).

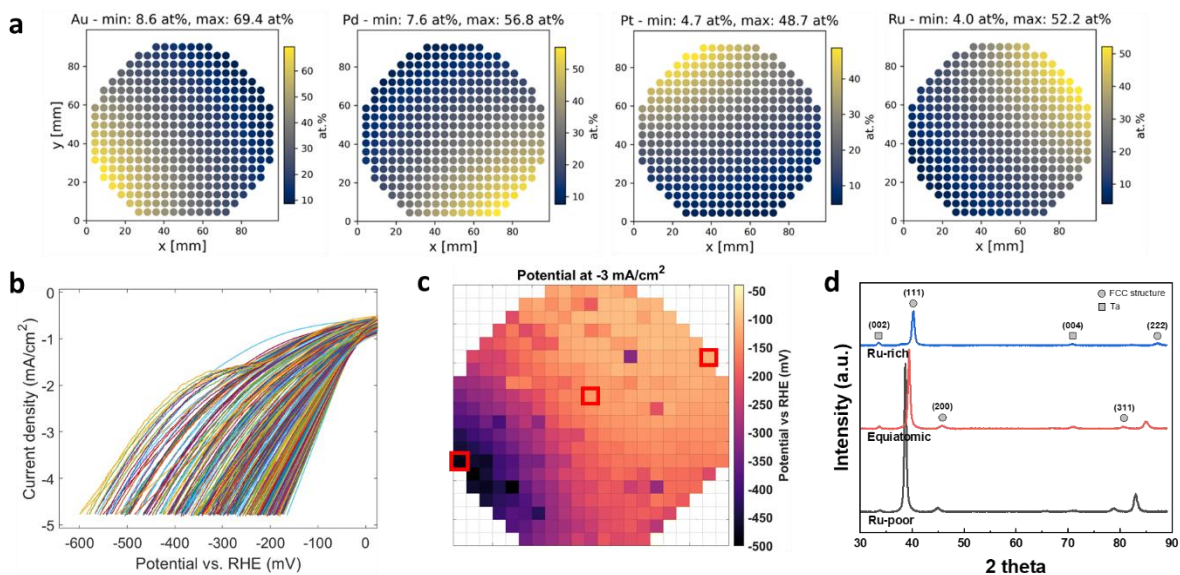

**Figure 1.** High-throughput screening data a) Color-coded composition maps of Au, Pd, Pt, and Ru in the quaternary Au-Pd-Pt-Ru CCSS thin-film materials library. b) Linear sweep voltammograms acquired using a high-throughput scanning droplet cell in the hydrogen evolution region. c) Electrocatalytic activity map (overpotential at a current density of -3 mA cm$^{-2}$) with the preselected MAs for further TEM analysis marked in red squares. d) Corresponding X-ray diffraction (XRD) patterns of the three MAs of the thin-film library.

XRD patterns of the three thin films show face-centered cubic (FCC) structures (solid circles), with no evidence of hexagonal close-packed (HCP) phases (**Figure 1d**). A diffraction peak from the Ta adhesion layer (solid square) between the films and the sapphire (0001) substrate is observed. With increasing Ru content, the diffraction peaks shift to higher angles, indicating lattice contraction. Although pure Ru is thermodynamically stable in the HCP phase, it is stabilized in the FCC structure of the quaternary solid solution[26]. Given that FCC Ru (~0.380 nm)[27] has a smaller lattice parameter than FCC Au (~0.407 nm), the observed peak shift is consistent with Ru alloying. Lattice parameters, calculated from the (111) and (222) reflections using Bragg's law, are ~0.403 nm (Ru-poor), ~0.395 nm (equiatomic), and ~0.388 nm (Ru-rich), consistent with the lattice parameters of the constituent elements. The use of (111) and (222) reflections ensured accuracy, given the strong (111) texture particularly in the Ru-rich film. These results show that lattice contraction upon Ru addition is coupled with compositional complexity, where substantial atomic size mismatch – particularly between Ru and Au – generates lattice strain that either accumulates or relaxes locally.

Cross-sectional TEM bright field images (**Figure 2**) reveal columnar grains with film thicknesses of ~145–185 nm. A Ta adhesion layer of ~25–30 nm is observed below the films, enhancing bonding of the CCSS to the sapphire substrate. The reduction in film thickness with increasing Ru content originates from Ru's lower sputter yield compared to Au, Pt, and Pd, leading to a reduced net deposition rate at the Ru-rich area in the library[28]. Average grain sizes decrease with increasing Ru content: ~30.2 ± 3.3 nm (Ru-poor), ~22.7 ± 2.0 nm (equiatomic), and ~16.2 ± 2.2 nm (Ru-rich). Planar defects which relaxes the lattice strain are visible in all three films. These planar defects are largely parallel to the substrate, but the contrast and number density differ, suggesting distinct defect morphologies across compositions.

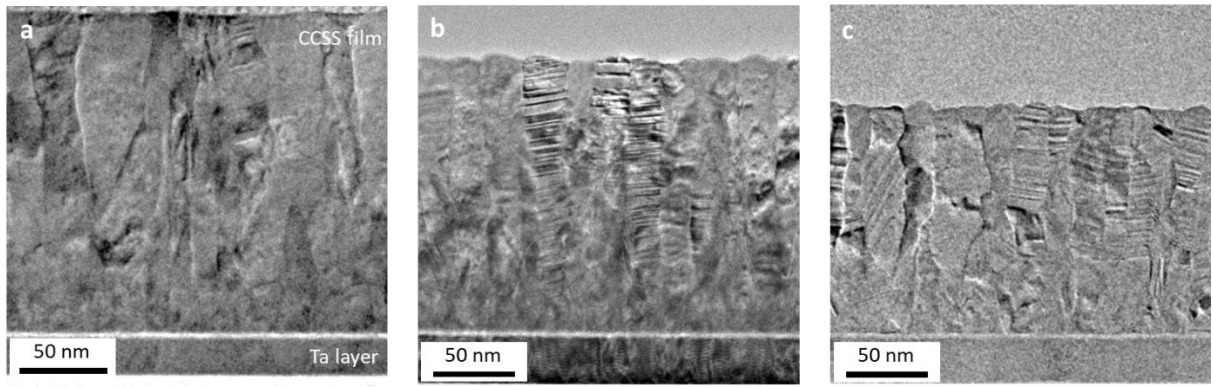

**Figure 2.** Grain morphology of three representative Au-Pd-Pt-Ru CCSS thin films acquired in cross-sections using TEM images a) the Ru-poor, b) the equiatomic, and c) the Ru-rich thin films.

To investigate these planar defects in detail, high magnification TEM images were obtained **(Figure 3)**. The Ru-poor film contains relatively few planar defects which are separated by 7–20 nm, mostly aligned parallel to the substrate, with occasional oblique orientations. The equiatomic film exhibits a higher density of planar defects (1-10 nm wide), oriented predominantly perpendicular to the growth direction. In contrast, the Ru-rich film shows dense arrays of planar defects with very small spacing that individual boundaries are difficult to resolve, again oriented largely perpendicular to the growth direction.

High-resolution TEM (HRTEM) images **(Figures 3d-f)** further elucidate the nature of these defects. In the Ru-poor film **(Figure 3d)**, the fast Fourier transformation (FFT) patterns acquired along the [001] zone axis are categorized by dashed orange lines for the matrix and dashed red lines for the twin, showing mirror symmetry across the twin boundary (TB) and identifies the defects as nanotwins. The TB is marked by white dashed lines in the TEM image, and the annotated lattice orientations of the matrix, twin region, and the TB correspond to the indexing in the FFT patterns. In the equiatomic film **(Figure 3e)**, similar FFT patterns confirm that the planar defects are also nanotwins; however, at the same magnification a higher density and smaller spacing between TBs are evident compared with **Figure 3d**. These twins are well aligned parallel to the substrate and exhibit growth along the {111} planes, consistent with the close-packed orientation of the FCC lattice.

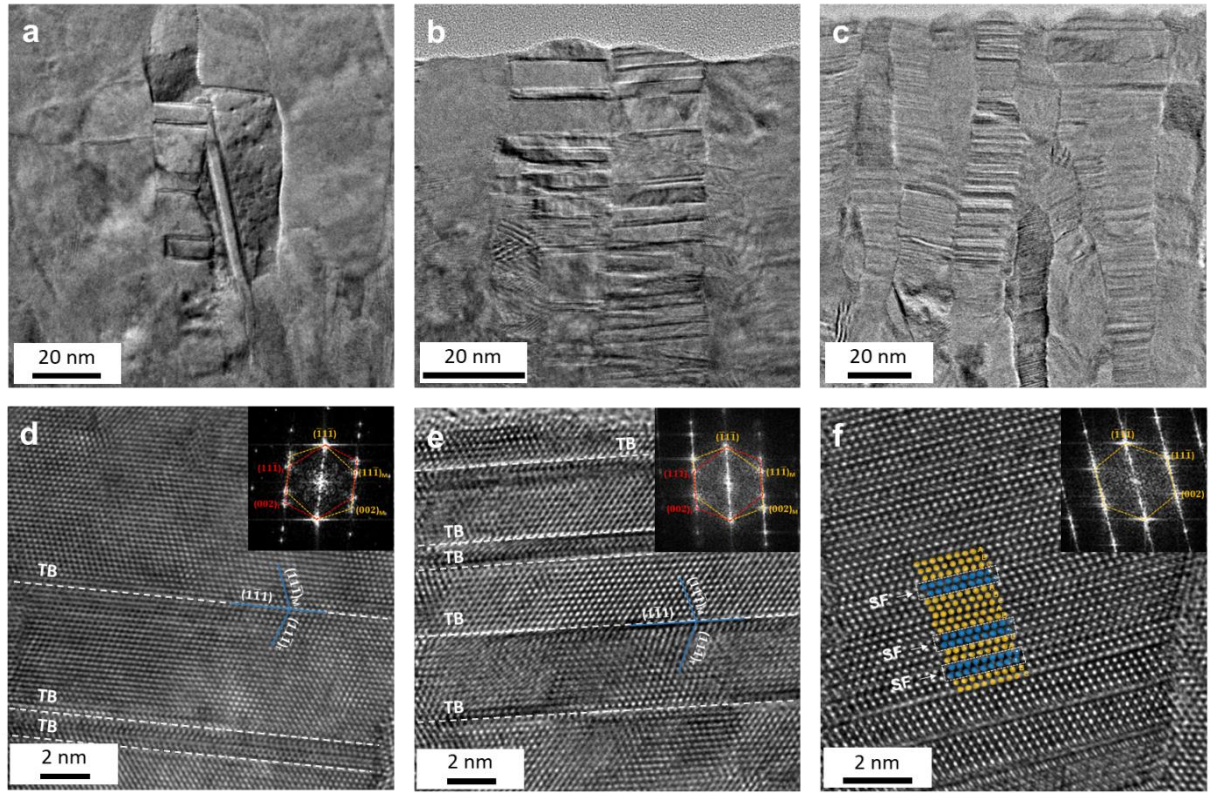

**Figure 3.** Planar defect formation of a), d) the Ru-poor, b), e) the equiatomic, and c), f) the Ru-rich thin films. d)-f) are magnified HRTEM images, the insets show FFT patterns of twins (d) and e)) and matrix, indicated by orange and red lines, respectively.

In contrast, the Ru-rich film **(Figure 3f)** shows extensive stacking faults (SFs) rather than nanotwins. The FFT pattern along the [011] zone axis displays streaking in addition to the diffraction spots, characteristic of SFs. To visualize the local stacking sequence, atomic columns in the matrix and within SF regions are marked by orange and blue solid circles, respectively; the accompanying sequence labels illustrate that the atomic sequence of FCC with ABCABC… order is locally disrupted by missing one {111} layer (e.g., ABCA(B)CAB…), which is a characteristic of an intrinsic stacking fault. The high SF density indicates a local tendency toward HCP-like ordering (ABABAB…), in line with Ru's thermodynamic preference for the HCP structure. As the Ru content increases, planar defect formation becomes more dominant, enabling partial structural relaxation towards HCP-like stacking.

With alloying more Ru in the CCSS film, it is also necessary to investigate possible crystallographic orientation changes in addition to the planar defect formation. The orientation maps **(Figure 4)** acquired on plan-view samples using 4D-STEM with a pixelated detector shows out-of-plane **(Figures 4a and 4e)** and in-plane **(Figures 4b-c and 4f-g)** orientations of the equiatomic and the Ru-rich films. The 4D-STEM analysis for the Ru-poor film was skipped as the others showed a much higher electrochemical activity. The out-of-plane orientation maps of both thin films reveals a strong (111) texture, described by blue colors according to the legend, consistent with the XRD results. The strong (111) orientations reflect the low surface energy of FCC {111} planes, favoring growth along this direction during film deposition. However, the in-plane orientation maps with sample orientations of (001) and (010) show random orientation distributions of both films. The image quality maps **(Figures 4d and 4h)** highlight grain boundaries. The grain sizes are consistent with cross-sectional TEM images.

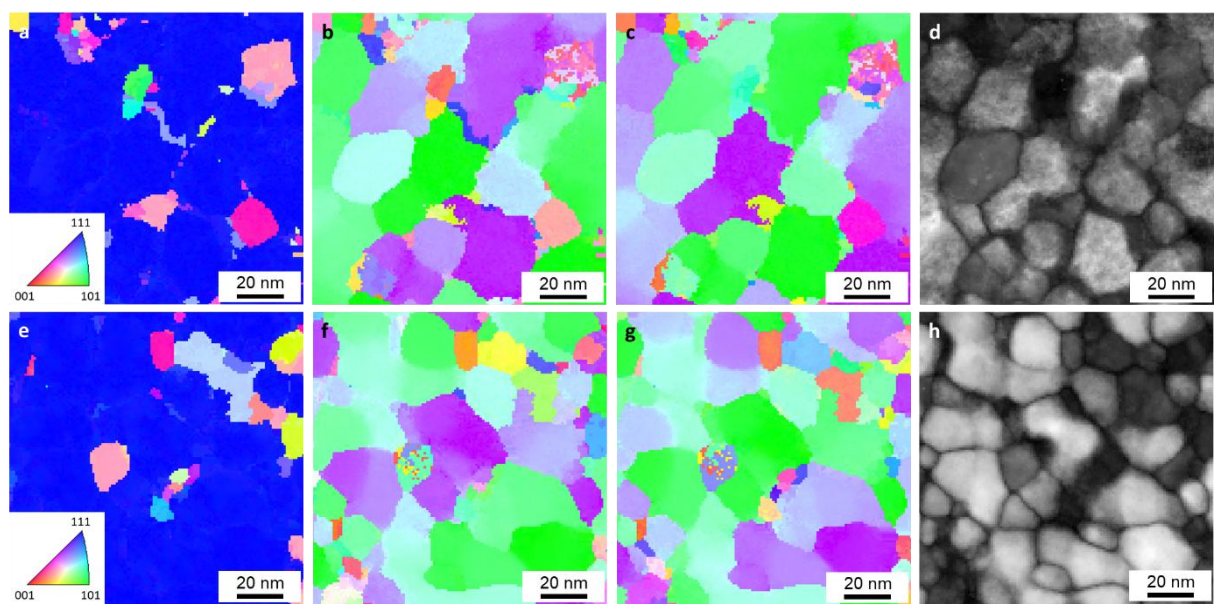

**Figure 4.** 4D-STEM orientation maps of plan-view samples of a)-d) the equiatomic and e)-h) the Ru-rich sample. a), e) out-of-plane orientation maps, b), f) in-plane (100) orientation maps, c), g) in-plane (010) orientation maps, and d), h) image quality maps of the equiatomic and the Ru-rich thin films.

In addition to the grain interior, the APT **(Figure 5)** was used to investigate the local atomic structure near defects, including grain boundaries and planar defects in the equiatomic and the Ru-rich films. **Figures 5a and 5d** reveal locally inhomogeneous elemental distribution, where Au, Pd, Pt, and Ru are marked in orange, red, magenta, and blue, respectively. Whereas the elements appear approximately homogeneous within grain interiors, chemical enrichment at grain boundaries is observed in both films, appearing much clearer in the Ru-rich film than in the equiatomic film. Although local compositional fluctuations may form around TBs and SFs, these defects are not revealed in APT, which could be due to their high density, as shown by STEM **(Figures 2 and 3)**. This leads to strong ion trajectory aberrations[29,30] that degrade the spatial resolution of APT, causing smearing of the enrichment and resulting in element mixing around TBs and SFs. The absence of elemental enrichment at TBs and SFs could be due to the low energy of these defects, particularly when coherent TBs are present.

It is known that as-deposited CCSS produced by sputtering exhibit a nanocrystalline structure with grains elongated along the film growth direction[31,32]. A similar scenario is observed in the current alloys. Cross-sectional images of our CCSS films **(Figure 2)** reveal the elongated grains in growth direction which are equiaxed as shown in the plan-view images where grain boundaries are clearly visible **(Figure 4, and Figures 5b and 5e)**. Because the APT sample axis is parallel to the film growth direction and the spacings between the modulations match the grain sizes, the chemically modulated regions **(Figures 5c and 5f)** correspond to grain boundaries. The black rectangle indicates the selected region of interest (ROI) across grain boundaries (GBs) for quantitative composition analysis. The red arrows mark the positions of two GBs. The grey bands in **Figures 5c and 5f** mark the corresponding positions of the GBs indicated in **Figures 5a and 5d**, respectively. Compositional analysis with magnified regions shows element-specific enrichment and depletion at these boundaries. In the equiatomic and Ru-rich films, Au and Pd are enriched at grain boundaries relative to their contents in grain interiors, although the amounts differ. In the equiatomic film, line scans show depletion of Pt and Ru at grain boundaries, whereas in the Ru-rich film only Ru appears depleted. Notably, the Ru depletion at grain boundaries is greater in the Ru-rich than in the equiatomic film, indicating that the grain interiors of the Ru-rich film contain more Ru than the nominal composition measured by SEM-EDX. Similarly, the Au and Pd contents within the grains are lower than their nominal compositions in both films. In the region where the grain boundaries terminate at the surface, different sites for the electrochemical reaction could evolve.

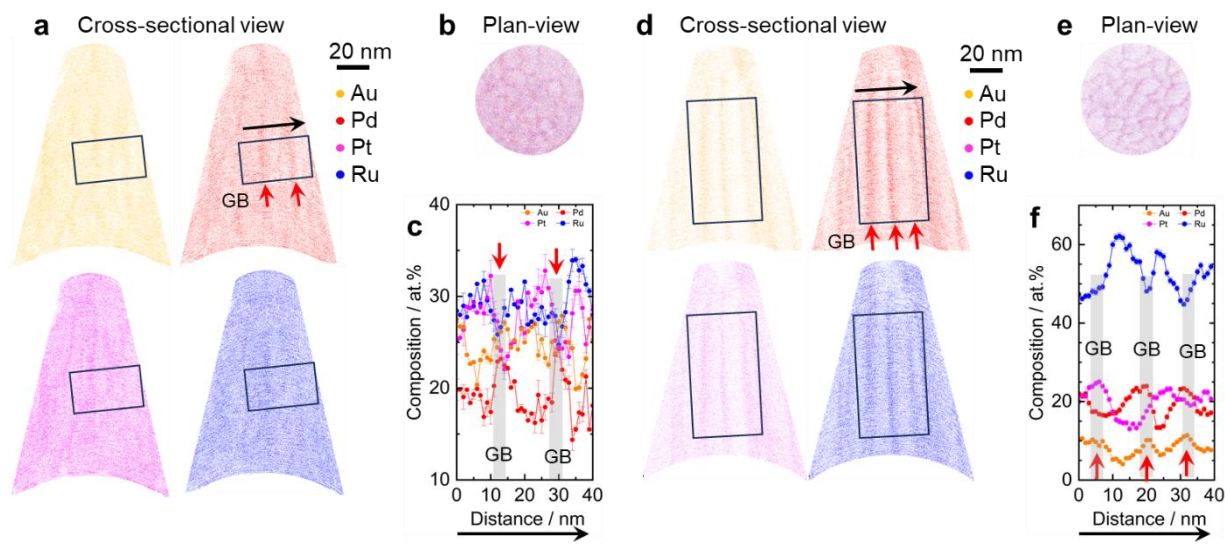

**Figure 5.** Elemental distributions of Au, Pd, Pt, and Ru in a) the equiatomic and d) the Ru-rich viewed from the cross-section of the APT needle-shaped sample. Plan-view of the atom maps for b) the equiatomic and e) the Ru-rich. One-dimensional (1D) composition profiles of c) the equiatomic and f) the Ru-rich for the selected ROI shown in a, plotted along the black arrow direction indicated in the Pd map.

The observed enrichment (Au and Pd) and depletion (Ru) at grain boundaries suggests that a higher content of Ru atoms in the grains is associated with stacking fault formation. This is consistent with a thermodynamic preference of Ru to form the HCP phase. Au, Pd, and Pt have FCC structure but exhibit different stacking fault energies; Pd and Pt have higher stacking fault energies than Au[33], implying that higher Pd and Pt contents generally prevent the formation of twins and stacking faults, as twin formation energies scale with stacking fault energy. Given the high grain boundary density in these CCSS films, grain boundaries act as preferential segregation sites that drive local composition fluctuations in their vicinity. Furthermore, since the grain boundary energies of the constituents are larger than their stacking fault energies[34], excess atoms would preferentially be enriched at grain boundaries rather than to the coherent interfaces of planar defects, and vice versa. This atomic configurations at defect structures could be governed by the formation energy of defects, which is also relevant to strain relaxation. The Ru atoms within the grains that are directly exposed to the (sub)surfaces could largely influence the electrocatalytic activities of the CCSS thin films. While compositional effects dominate in CCSS electrocatalysts, microstructural complexity may still play a supporting role in the electrocatalytic activity as defects terminating at the surface might change the active sites. The role of individual planar defects – if terminating at the surfaces - will be a focus of our future work.

## 3. Conclusion

This study investigates the effect of Ru content on the electrocatalytic activity of the HER and planar defect formation of Au-Pd-Pt-Ru CCSS thin films. To obtain detailed structural information, three representative thin films were selected from a thin-film materials library which was investigated for hydrogen evolution in alkaline media. With increasing Ru content, the lattice parameters decrease, while the strong (111) texture remains. Whereas the three films across the library show a single solid solution phase, various defect structures including grain boundaries and planar defects are observed in the CCSS thin films. The grain boundaries give rise to local atomic fluctuations, altering compositional distributions at an atomic scale and consequently influencing electrochemical activity at these sites. Notably, high densities of nanotwins and SFs evolve in the equiatomic and the Ru-rich film, respectively. These planar defects exhibit coherent boundaries with the matrix, so enrichment at planar defects is scarcely observed. By contrast, specific elements are enriched or depleted at grain boundaries owing to differences in the formation energies of grain boundaries and planar defects. This study provides a guidance for the rational design of CCSS thin film electrocatalysts by considering both compositional and microstructural complexity.

## 4. Experimental Section
### Synthesis of Au-Pd-Pt-Ru CCSS Thin-Film Materials Library

The Au-Pd-Pt-Ru thin-film materials library (ML) was synthesized using a four-cathode magnetron co-sputter system (Polaris, AJA International) using four 3.81 cm (1.5 inch) targets simultaneously (Au (99.99 %, Sindlhauser Materials), Pd (99.9%, Sindlhauser Materials), Pt (99.99 %, ESG Edelmetallservice), and Ru (99.99 %, Evochem advanced Materials). A Ta (99.95 %, Sindlhauser Materials) adhesion layer was sputtered onto a 10 cm diameter single-side polished sapphire substrate (SITUS Technicals, c-plane orientation) and moved under atmosphere condition to the co-sputter system. All targets were pre-cleaned for 120 s at 1.33 Pa and 20 W for DC and 30 W for RF power supplied targets. The deposition was carried out at room temperature with a deposition pressure of 0.5 Pa (base pressure $9.7 \times 10^{-5}$ Pa) in Ar (99.9999 %) atmosphere. All cathodes are equally distributed around the static substrate with a target-to-substrate distance of 10 cm. The positioning of each deposition source with respect to the substrate surface leads to wedge-shaped deposition profiles. The overlay of these profiles from the four sources forms the continuous composition gradients of the ML. Based on deposition rate the deposition time was chosen to be 900 s to get an estimated layer thickness of 150 nm at the center of the ML. All other sputter parameters are presented in Table 1.

Table 1 Specific sputter parameter for the synthesis of the ML.

| Targets | Au | Pd | Pt | Ru |
|---|---|---|---|---|
| Power Supply | RF | DC | DC | DC |
| Deposition Power (W) | 30 | 16 | 16 | 29 |
| Deposition rate (nm/s) | 0.00576 | 0.00333 | 0.00362 | 0.00174 |

**High-throughput approach (screening)**

The ML was investigated using a high-throughput approach for structural, compositional, and functional characterization. The chemical compositions of the ML were characterized automatically by measuring 342 measurement areas (MAs) (4.5 mm x 4.5 mm) using energy-dispersive X-ray spectroscopy (EDX) in a tungsten sourced scanning electron microscope (JSM-5800 LV, JEOL) equipped with an EDX detector (INCA X-act, Oxford Instruments) with a nominal measurement accuracy of ±1 at%. Before the characterization of the whole ML, a calibration on a Co standard was performed. Each MA was acquired for 60 s with an acceleration voltage of 20 kV at a magnification of 600. The quantification process was executed using *Oxford Instruments' INCA* software, yielding normalized compositions for each MA. Following the exclusion of the sapphire substrate signal, the analysis indicated that the ML center exhibited a composition of $Au_{30}Pd_{28}Pt_{19}Ru_{23}$.

The crystal structures of the ML were examined by X-ray diffraction (XRD) using a diffractometer (D8 Discover, Bruker) equipped with a Cu X-ray source (Incoatec IμS High-Brilliance Microfocus, 0.15418 nm) and a 2D detector (Vantec-500). Each MA 1D diffraction pattern was obtained by converting three merged 2D diffraction frames measured at 2θ positions of 30°, 50°, and 70°, covering a 2θ range of 20 - 90°. Conversion and background subtraction were performed using Bruker's *DIFFRAC.EVA* software.

**Electrochemical measurements**

All electrochemical measurements were conducted using a high-throughput SDC setup, which automatically evaluates 342 MAs of the thin-film materials library. The measurements were conducted in a conventional three-electrode configuration using a Pt wire as counter electrode and Ag|AgCl|3M KCl reference electrode. The counter and the reference electrodes were placed inside the SDC tip made of polytetrafluoroethylene (PTFE) and filled with 0.05 M KOH electrolyte. Before each new measurement starts, the electrolyte is automatically replaced. By pressing the tip on the surface of the sample, an electrochemical cell is formed between the investigated MA (working electrode) and the electrolyte droplet. The circular area of the working electrode (0.00735 cm$^2$) is defined by the tip opening (1 mm). The applied pressing force is monitored with a force sensor integrated into the tip holder. The measured current is normalized by the geometric surface area. All potentials are calculated and reported versus the RHE through the equation below where $E_{(Ag|AgCl|3M\ KCl)}$ is the potential measured versus Ag|AgCl|3M KCl; 0.210 V is the standard potential of the Ag|AgCl|3M KCl reference electrode at 298 K; 0.059 is the result of $(RT)/(nF)$, where R is the gas constant, *T* the temperature (298 K), *F* the Faraday constant, and *n* the number of transferred electrons during the reaction. The Nernst equation was used to calculate the values:

$$E_{RHE} = E_{(Ag|AgCl|3M\ KCl)} + 0.210 + 0.059\ pH.$$

**High-resolution (scanning) transmission electron microscopy**

High-resolution transmission electron microscopy (HR-TEM) was carried out using an image Cs-corrected Titan Themis 80-300 from Thermo Fisher operated at 300 kV, while high angle annular dark field – scanning transmission electron microscopy (HAADF-STEM) with EDX analysis was conducted using a Thermo Fisher Titan Themis microscope with probe Cs correction at an accelerating voltage of 300 kV. STEM–EDX data were acquired using a Super-X G1 system with four detectors at a tilted geometry. The quantified compositions were normalized to 100%. Signals from C, O, Cu, and Ga were excluded from the quantification. The four-dimensional-STEM (4D-STEM) was utilized for acquiring grain orientation maps. The data was collected using the electron microscope pixel array detector (EMPAD) equipped with the Cs-probe corrected Titan Themis microscope. For the acquisition, a cameral length of 1.5 m and a probe convergence semiangle of 0.65 mrad were employed. The 4D-STEM data was reconstructed using the commercial software (automated crystal orientation mapping (ACOM); Nanomegas) with an input of crystal structure information of specimens. Cross-sectioned and plan-view TEM samples were prepared using a focused ion beam (FIB) system (FEI Helios G4 CX) operated at 30 kV. Low energy cleaning was also applied to each side of the TEM sample in the final step to remove beam damage. The cross-sectioned TEM lamella was used for investigating overview of microstructure in thin films, while the plan-view TEM samples was used for acquiring the 4D-STEM dataset for the orientation mapping.

**Atom probe tomography reconstruction**

APT specimens were extracted parallel to the film growth direction using the same FIB (FEI Helios G4 CX), as mentioned above for TEM specimen preparation. Protection layer of about 20 nm thick carbon followed by a 200 nm thick Pt were deposited using an e-beam on the selected area. APT analysis was performed using laser pulses on a local electrode atom probe (LEAP 5000XR, Cameca Instruments) at 65 K with 50 pJ laser energy at a pulse repetition rate of 125 kHz and a detection rate of 0.005 atoms per pulse. The APT data were reconstructed and analyzed using the AP suite 6.3 software.

**Supporting Information**

**Acknowledgements**

The authors acknowledge the Deutsche Forschungsgemeinschaft (DFG, German Research Foundation) in the – SFB 1625, project number 506711657, subproject B03 (M. Joo, H. Xiu, C, Somsen, and C. Scheu), subproject Z and A01 (S. Baha and A. Ludwig), subproject B01 (Y. Li), subproject C01 (R. Zerdoumi and W. Schuhmann), and S (A. Kostka). We thank Ibrahim Akinci for fabrication of the materials library as part of his student project.

**Conflict of Interest**

The authors declare no conflict of interest.

# Influence of Ru content on electrocatalytic activity and defect formation of Au-Pd-Pt-Ru compositionally complex solid solution thin films

Miran Joo[1], Huixin Xiu[1,2], Sabrina Baha[3], Ridha Zerdoumi[3,4], Ningyan Cheng[1], Christoph Somsen[5], Yujiao Li[6], Aleksander Kostka[6], Wolfgang Schuhmann[4], Alfred Ludwig[3,6] and Christina Scheu[1*]

ToC figure

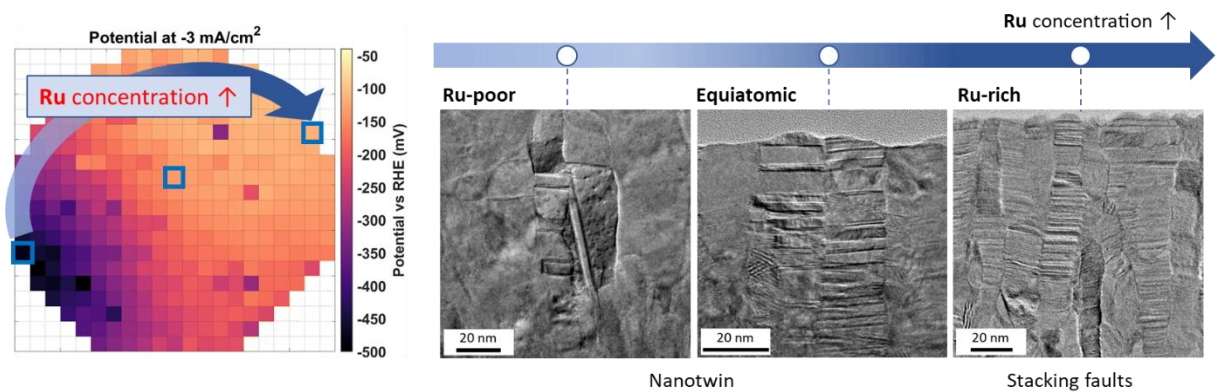

Transmission electron microscopy reveals that the formation of planar defects in Au–Pd–Pt–Ru CCSS thin films depends on Ru content, which in turn influences HER electrocatalytic activity. As the Ru content increases, the density of nanotwins initially rises; with further Ru addition, stacking faults form. These defects may diversify surface atomic arrangements in the CCSS. Overall, these results highlight how microstructural complexity can impact electrocatalytic performance and broaden our understanding of microstructural design strategies for CCSS electrocatalysts.

# Supporting Information

# Influence of Ru content on electrocatalytic activity and defect formation of Au-Pd-Pt-Ru compositionally complex solid solution thin films

Miran Joo[1], Huixin Xiu[1,2], Sabrina Baha[3], Ridha Zerdoumi[3,4], Ningyan Cheng[1], Christoph Somsen[5], Yujiao Li[6], Aleksander Kostka[6], Wolfgang Schuhmann[4], Alfred Ludwig[3,6] and Christina Scheu[1*]

[1] Nanoanalytics and Interfaces, Max Planck Institute for Sustainable Materials, 40237 Düsseldorf, Germany

[2] School of Materials and Chemistry, University of Shanghai for Science and Technology, 516 Jungong Road, Shanghai, 200093, China

[3] Chair for Materials Discovery and Interfaces, Institute for Materials, Ruhr University Bochum, 44801 Bochum, Germany

[4] Analytical Chemistry-Center for Electrochemical Sciences (CES), Faculty of Chemistry and Biochemistry, Ruhr University Bochum, 44801 Bochum, Germany

[5] Chair for Materials Science and Engineering, Institute for Materials, Ruhr University Bochum, 44801 Bochum, Germany

[6] Center for Interface-Dominated High Performance Materials (ZGH), Ruhr University Bochum, 44801 Bochum, Germany

*Corresponding author

E-mail address: c.scheu@mpi-susmat.de

| at.% | SEM-EDX | | | STEM-EDX | | |
|---|---|---|---|---|---|---|
| | Ru-poor | Equiatomic | Ru-rich | Ru-poor | Equiatomic | Ru-rich |
| **Au** | 67,9 | 26,8 | 8,70 | 63,2 | 26,8 | 8,60 |
| **Pd** | 13,0 | 23,3 | 21,5 | 14,5 | 17,7 | 18,9 |
| **Pt** | 14,9 | 23,6 | 17,7 | 16,7 | 31,1 | 21,1 |
| **Ru** | 4,20 | 26,3 | 52,1 | 5,60 | 24,4 | 51,4 |

**Table S1.** Results of compositional analysis for the Ru-poor, the equiatomic, and the Ru-rich thin films, using SEM-EDX and STEM-EDX techniques.